\documentclass{sigchi}

\toappear{A version of this article was featured in the  Interactions magazine of the ACM, volume 15, issue 2 (March 2008), pages 50-53. http://doi.acm.org/10.1145/1340961.1340974 }
\usepackage{balance}  % to better equalize the last page
\usepackage{graphics} % for EPS, load graphicx instead
\usepackage{times}    % comment if you want LaTeX's default font
\usepackage{url}      % llt: nicely formatted URLs

\makeatletter
\def\url@leostyle{%
  \@ifundefined{selectfont}{\def\UrlFont{\sf}}{\def\UrlFont{\small\bf\ttfamily}}}
\makeatother
\def\pprw{8.5in}
\def\pprh{11in}

\usepackage[pdftex]{hyperref}
\hypersetup{
pdftitle={Empowering Kids to Create and Share Programmable Media},
pdfauthor={LaTeX},
pdfkeywords={creative learning, children, online communities, end-user programming, social computing},
bookmarksnumbered,
pdfstartview={FitH},
colorlinks,
citecolor=black,
filecolor=black,
linkcolor=black,
urlcolor=black,
breaklinks=true,
}

\begin{document}

\title{Empowering Kids to Create and Share Programmable Media}

\numberofauthors{3}
\author{
  \alignauthor Andr\'es Monroy-Hern\'andez\\
    \affaddr{MIT Media Laboratory}\\
    \email{andresmh@media.mit.edu}\\
  \alignauthor Mitchel Resnick\\
    \affaddr{MIT Media Laboratory}\\
    \email{mres@media.mit.edu}\\
}

\maketitle

\section{Abstract}
This article reflects on the first eight months of existence of the Scratch Online Community by discussing the design rationale and learning theories underlying Scratch and its website.


%\category{H.5.m.}{Information Interfaces and Presentation (e.g. HCI)}{Miscellaneous}

\section{Introduction}
There are now many websites, such as Flickr\footnote{\url{http://flickr.com}} and YouTube\footnote{\url{http://youtube.com}} and blogs, which support user-generated content, enabling people to create and share text, graphics, photos, and videos. But for the most part, Web 2.0 does not include interactive content. People interact with Web-based animations and games all the time, but few people can create and share their own interactive content.

The Scratch project\cite{resnick_sowing_2007} from the MIT Media Lab aims to change that, making it easy for everyone, especially children and teens, to create and share interactive stories, games, and animations on the Web, in the participatory spirit of Web 2.0. With the Scratch programming environment\cite{resnick2005networked}, users snap together graphical programming blocks to control the actions and interactions of rich media content, including photos, graphics, music, and sound. Then they upload their interactive creations to the shared Scratch website\footnote{\url{http://scratch.mit.edu}}, where other members of the Scratch community can interact with the projects on the site and download the original source code to examine or modify the project\cite{monroy2007scratchr}.

\begin{figure}
\begin{center}
\includegraphics[width=3.4in]{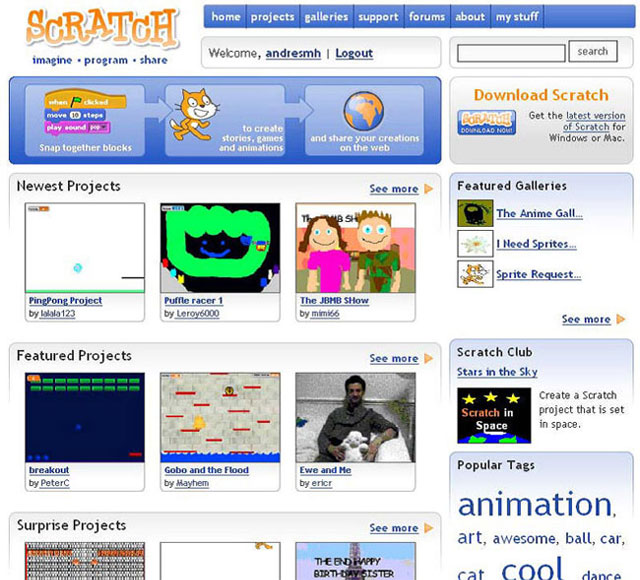}
\caption{The home page of the the Scratch website highlights projects contributed by the user community.}
\label{fig:homepage}
\end{center}
\end{figure}

\begin{figure}
\begin{center}
\includegraphics[width=3.4in]{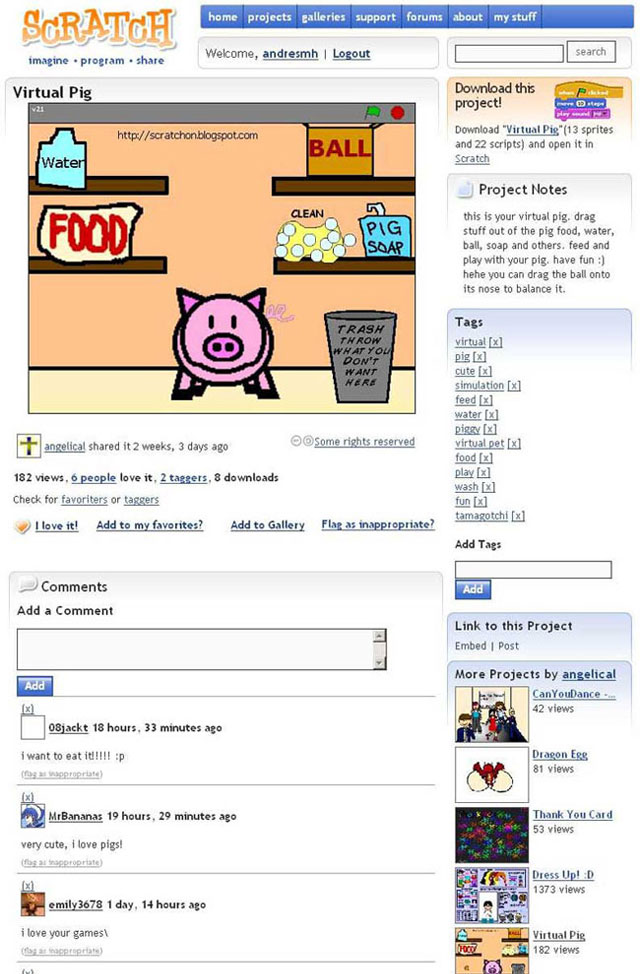}
\caption{A project page on the Scratch website allows people to interact  projects. }
\label{fig:project}
\end{center}
\end{figure}

The Scratch website (see Figures \ref{fig:homepage} and \ref{fig:project}) offers an alternate model for how children might use the Web as a platform for learning, enabling them to create and share personally meaningful projects, not simply access information. Children create and share Scratch projects as a way to express themselves creatively, much as they would paint a picture or build a castle with LEGO bricks. In the process they not only learn important math and computer science concepts, but they also develop important learning skills: creative thinking, effective communication, critical analysis, systematic experimentation, iterative design, and continual learning. We believe that the ability to produce (not simply interact with) interactive content is a key ingredient to achieving digital literacy and becoming a full participant in the interactive online world.

\section{Learning Through Online Community}
The Scratch Online Community see makes programming more engaging by turning it into a social activity. Hobbit, a 14-year-old member of the community explains: \textit{``When I think about it, recognition for my work is what really drew me into Scratch. Other things played a part, but the feeling that my work would be seen is what really motivated me.''} The website provides a wide range of entry points for community interactions. Children comment on projects, upload their own projects, and can become involved in existing projects. The site is also a repository of user-generated content that serves as a source of inspiration and appropriable objects for new ideas. Users can connect with each other, forming a social network of creators and collaborators through the use of ``friendships,'' galleries (groups of projects based on a topic), and forums where users can post their questions or interests to be discussed with others.

Inspired by Jenkins's description of the states of participation in fan-fiction communities \cite{jenkins2006convergence}, we put forward the idea that members of user-generated-content communities tend to move in four different roles or states of participation: passive consumption, active consumption, passive production, and active production. In order to build a successful community, it is essential for the sites in question to support and welcome users regardless of which state of participation they fall into. For example, Lave and Wenger argue that ``peripheral participation'' is a legitimate form of engagement\cite{lave1991situated}. These roles/states are the core of most user-generated-content sites, and the Scratch community addresses them in a relevant way for the specific audience and type of content.

\begin{description}

\item[Passive consumer.] Online communities often refer to these people as lurkers. In this state, people assess the community to understand their values and ideas. In the case of Scratch, this involves the act of browsing the different categories and interacting with Scratch projects that other people have created. While this is the most passive state, the passive consumer alters the system simply by viewing because the number of views is counted and presented publicly.

\item[Active consumer.] An active consumer participates in the community by providing metadata. Active Scratch consumers contribute their ideas by commenting, tagging, and rating projects.

\item[Passive producer.] In this state, users create projects, sometimes inspired by other projects they have seen in the community, but do not necessarily feel compelled or ready to share them to the community. 
\item[Active producer.] An active producer not only consumes but also contributes to the repository of projects. This person gives feedback to other people's projects, gets inspired, and also provides inspiration. An analysis of the usage of the website showed that the number of projects a user creates is correlated with the level of activity by that user on projects created by others. That is, there is a correlation between the number of projects a user creates and the number of a) comments posted on other people's projects, b) views on others' projects, c) projects marked as favorites, d) projects marked as ``I love it!,'' and e) projects downloaded. Smaller correlations were found in regard to tags. Other people often recognize these active producers' level of involvement. Members in this state feel invested in the community—it is one of the most important assets of the Scratch online community.
\end{description}

\section{Sharing and Collaboration}

\begin{figure}
\begin{center}
\includegraphics[width=3.4in]{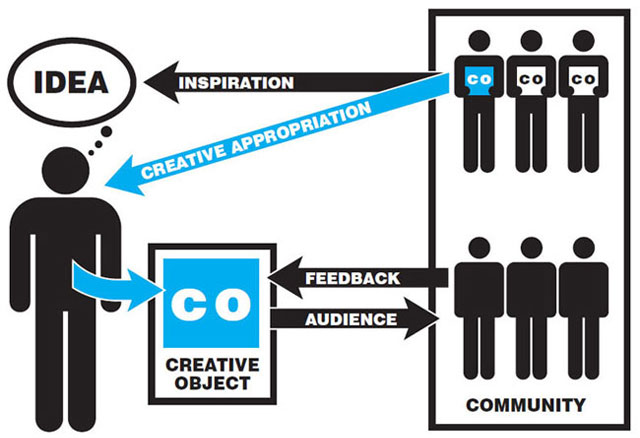}
\caption{Scratch users build on one another's projects 
through ``creative appropriation.''}
\label{fig:appropriation}
\end{center}
\end{figure}

We use the term \emph{creative appropriation} to refer to the utilization of someone else's creative work in the making of a new one (see Figure \ref{fig:appropriation}). Professional programmers are very familiar with this concept, as a great deal of their work is based on programs and algorithms created by others. With Scratch, we wanted to introduce children and teens to this approach, because learning in the context of a community is not only more convenient, but is also more rewarding and engaging.

\begin{figure}
\begin{center}
\includegraphics[width=3.4in]{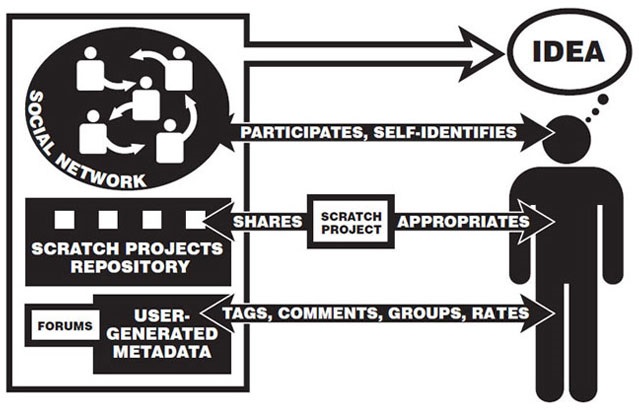}
\caption{Scratch users contribute to (and learn from) the online community in many ways.}
\label{fig:activities}
\end{center}
\end{figure}

One of the main goals of the Scratch online community is to foster the idea of learning from each other by building on other people's ideas or projects (see Figure \ref{fig:activities}). This is one of the reasons why it is always possible for a member of the community to download the source code of any project. Additionally, users of the community often create their projects after being inspired by other projects they see. In this type of creative appropriation, no code or media is reused; instead, it is the idea or concept that is appropriated to create a new project. This type of appropriation often leads to the emergence of trends in the community. One of these trends was started by an interactive ``dress up'' project created by an 11-year-old girl from South Africa. The project was a digital version of a traditional paper doll: The viewer could choose the skin color, hair, and clothing of the doll. Projects tagged as ``dress up'' are so popular that they often go to the ``Top Viewed'' section of the front page with hundreds if not thousands of views. To date, there are more than 150 projects tagged as ``dress up.'' Ranging from a project about dressing up a hero to dressing up a famous TV star and original characters, ``dress up'' projects are as diverse as their creators.

The Scratch website serves as a repository of code and ideas that can be creatively appropriated to spawn new ideas and new projects. The Scratch website and the Scratch desktop environment make it very easy for this to happen. Fifteen percent of all of the 23,394 projects shared (as of August 14, 2007) were remixes of other projects. Of those, the types of changes made ranged from simple changes to images and sounds, to modifications of the actual programming code.

Every time a project gets shared on the Scratch website, the Scratch desktop application adds information about who shared the project and when. This information is used to automatically connect projects that are based on others. When a project is a remix of another, it displays a link to the original project, giving credit to the creator whose work has been remixed. Several members of the community have posted messages in the online forums expressing their concern about others ``copying'' their work. This controversy has provided an opportunity to discuss important ideas and differences between plagiarism and sharing.

\section{Mesh Inc.}

One of the early and ongoing collaborative efforts on the Scratch online community started when a 15-year-old girl from the U.K., screen name BeeBop, created a series of projects with animated sprites and encouraged others to use them in their projects.

\textit{``You can take any of these to use in your own project, or you can post a comment saying what you want and I can make it for you,''}
BeeBop explained. The same day, a 10-year-old girl, using the name MusicalMoon, wrote a comment saying that she liked BeeBop's animations and asking if BeeBop could create a project with \textit{``a mountain background from a bird's-eye view''} for use in one of her own projects. 
MusicalMoon also asked BeeBop to submit the project to Mesh Inc., a ``miniature company'' that MusicalMoon had created to produce ``top quality games'' in Scratch. MusicalMoon explained that \textit{``all you do is simply send in a project, I will review it back in the Mesh gallery, and, then, if it's good enough, I will grant you a member of Mesh Inc!''} 
MusicalMoon and BeeBop continued their exchanges and created an initial version of a collaborative project.

A few days later, Hobbit, the 14-year-old boy from New Jersey, discovered the Mesh Inc. gallery and offered his services: \textit{``I'm a fairly good programmer, and I could help with debugging and stuff.''} MusicalMoon asked Hobbit if he could solve a problem with a particular Mesh Inc. project: \textit{``I can't make characters jump so you're up.''} A day later Hobbit fixed the game and posted: \textit{``This is the new updated version, so now he can jump on the snow.''} MusicalMoon replied \textit{``gr8 job, Hobbit! I'll take this and carry on from here.''} Meanwhile, Hobbit decided to put his blogging skills to use and created a blog for Mesh Inc. where each of Mesh Inc.'s members is listed with their corresponding positions. MusicalMoon was selected as the ``chairlady.'' Later, an 11-year-old boy from Ireland calling himself Marty was added to the Mesh staff as the expert in ``scrolling backgrounds.''

As others witnessed these interactions happening, Mesh Inc. got a lot of recognition in the community and many people started to ``audition'' for Mesh Inc. BlueRiver, a 12-year-old girl from Russia, now leads the ``character design'' and ``sound operations'' along with GreenDinosaur, a 10-year-old boy from the U.S., who holds the title of ``story writer.'' Other Scratch community members, inspired by Mesh Inc, have created their own similar companies.

\section{Appendix: Usage Statistics}

The Scratch Online Community was beta released on March 4, 2007. The community started with only the 20 participants who were involved in a Scratch workshop. On the morning of May 14, 2007, the website was officially launched. Several news outlets and social news websites featured the Scratch website on their front pages. In a matter of hours the server and the website could not handle the traffic and the website went down several times.

As of December 9, 2007:
\begin{itemize}
\item the site has received 10,373,606 page views
\item there have been 1,708,857 sessions
\item the site was visited by 1,176,042 unique visitors
\item 56,352 projects have been shared
\item 915,489 scripts have been created
\item 317,142 sprites have been created
\item 53,639 members have registered
\item 10,743 individuals have contributed content
\item 181,230 comments have been posted on projects, galleries, and forums
\end{itemize}

While the majority of the users come from the United States, London is the city that generates the most number of visits. Visitors to the site come from 213 different countries, mainly from the U.S., U.K., Canada, Australia, Japan, Germany, Brazil, Spain, France, and India.

An analysis of usage data during the first five months showed that users are primarily age 8 to 17, with a peak at age 12. A good number of users are adult computer hobbyists and educators that create projects in Scratch, even though a lot of them know other professional programming languages. Some members of the community have emerged as mentors that help the beginners and provide advice.

Data also shows that age is not indicative of engagement. No correlation was found between age and number of projects ($r = 0.108, p < 0.001$). Also, surprisingly, no correlation was found between the number of posts on the text-based forums and age either ($r = -0.016, p = 0.007$). Even starting new threads on the forums is not correlated to age ($r = -0.016, p = 0.006$). Age was also not an indicator of the number of friends ($r = 0.065, p < 0.0001$).

While 70 percent of users are male, no statistically significant correlation was found between gender and the number of projects ($r = 0.001, p = 0.923$). This indicates that even though the majority of users are male, the females are as engaged in creating projects as the males. As we continue our work on Scratch, one main goal is to achieve broader participation across gender.

\section{Acknowledgements}

Scratch and the Scratch website have been developed by the Lifelong Kindergarten Group at the MIT Media Lab. The core development team includes: John Maloney, Natalie Rusk, Evelyn Eastmond, Tammy Stern, Amon Millner, Jay Silver, Han Xu, Eric Rosenbaum, Karen Brennan, Brian Silverman, and the authors. Special thanks to Ubong Ukoh, Kemie Guaida, Lis Sylvan, Chris Garrity, Lance Vikaros, and Chris Spence for their contributions to the Scratch website and online community. Yasmin Kafai, Kylie Peppler, Grace Chiu, and others at UCLA Graduate School of Education and Information Studies collaborated on the development of Scratch. All screen names in this article are pseudonyms. This material is based upon work supported by the National Science Foundation under Grant No. ITR-0325828. The Scratch project has also received financial support from the Intel Foundation, the LEGO Company, and MIT Media Lab consortia.

\section{About the authors}

\begin{description}
\item[Andr\'es Monroy-Hern\'andez,] Ph.D. student and  Samsung Fellow at the MIT 
Media Lab, has conceptualized and led the development of the Scratch online community.  Andr\'es is interested in the development of 
social software that fosters creative and  collaborative learning experiences. He has  worked in the software industry and at the 
Los Alamos National Laboratory. He  received a B.S. in electronic systems engineering from the Tecnol\'ogico de Monterrey 
in M\'exico
\item[Mitchel Resnick,] professor  of learning research at the 
MIT Media Lab, explores how new technologies can 
engage children and teens  in creative learning experiences. His Lifelong Kindergarten research group has developed many innovative educational technologies, including Scratch and the ``programmable bricks'' that were 
the basis for the LEGO MindStorms and PicoCricket robotics kits. Resnick cofounded the Computer Clubhouse project, an 
international network of after-school learning centers for youth from low-income communities, with more than 100 sites in 20  countries. Resnick earned a B.S. in physics from Princeton and an M.S. and Ph.D. in 
computer science from MIT. He is the author or coauthor of several books, including Turtles, Termites, and Traffic Jams.
\end{description}

% Balancing columns in a ref list is a bit of a pain because you
% either use a hack like flushend or balance, or manually insert
% a column break.  http://www.tex.ac.uk/cgi-bin/texfaq2html?label=balance
% multicols doesn't work because we're already in two-column mode,
% and flushend isn't awesome, so I choose balance.  See this
% for more info: http://cs.brown.edu/system/software/latex/doc/balance.pdf
%
% Note that in a perfect world balance wants to be in the first
% column of the last page.
%
% If balance doesn't work for you, you can remove that and
% hard-code a column break into the bbl file right before you
% submit:
%
% http://stackoverflow.com/questions/2149854/how-to-manually-equalize-columns-
% in-an-ieee-paper-if-using-bibtex
%
% Or, just remove \balance and give up on balancing the last page.
%
%\balance

\bibliographystyle{acm-sigchi}
\bibliography{interactions}
\end{document}